\documentclass[status=final,copyright,creativecommons]{eptcs} %

\usepackage{xspace}
\usepackage[svgnames,table]{xcolor}
\usepackage{amsmath}
\usepackage{amsfonts}
\usepackage{tikz}
\usetikzlibrary{calc,arrows.meta,shapes}
\usepackage{wrapfig}
\usepackage{paralist}
\usepackage{doi}
\usepackage{algorithm}
\usepackage{algpseudocodex}
\usepackage{cleveref}
\Crefname{section}{Sect.}{Sects.}
\Crefname{equation}{Formula}{Formulae}
\Crefname{lstlisting}{Listing}{Listings}

\DeclareMathOperator{\type}{\mathsf{type}}

\providecommand{\event}{FMAS 2021}

\title{Complete Test of Synthesised Safety Supervisors\\for Robots
  and Autonomous Systems\thanks{The second author is partially funded
    by the German Ministry of Economics, Grant Agreement~20X1908E.}}
\author{{Mario Gleirscher and Jan Peleska} \institute{Department of
    Mathematics \& Computer Science, University of Bremen, Germany}
  \email{\{gleirscher,peleska\}@uni-bremen.de}}

\def\titlerunning{Testing of Safety Supervisors}
\def\authorrunning{Mario Gleirscher and Jan Peleska}

\makeatletter\begin{document}
\maketitle

\begin{abstract}
  Verified controller synthesis uses world models that comprise all
  potential behaviours of humans, robots, further equipment, and the
  controller to be synthesised.  A world model enables quantitative
  risk assessment, for example, by stochastic model checking.  Such a
  model describes a range of controller behaviours some of
  which---when implemented correctly---guarantee that the overall risk
  in the actual world is acceptable, provided that the stochastic
  assumptions have been made to the safe side.  Synthesis then selects
  an acceptable-risk controller behaviour.  However, because of
  crossing abstraction, formalism, and tool boundaries, verified
  synthesis for robots and autonomous systems has to be accompanied by
  rigorous testing.  In general, standards and regulations for
  safety-critical systems require testing as a key element to obtain
  certification credit before entry into service.  This
  work-in-progress paper presents an approach to the complete testing
  of synthesised supervisory controllers that enforce safety
  properties in domains such as human-robot collaboration and
  autonomous driving.  Controller code is generated from the selected
  controller behaviour.  The code generator, however, is hard, if not
  infeasible, to verify in a formal and comprehensive way. Instead,
  utilising testing, an abstract test reference is generated, a
  symbolic finite state machine with simpler semantics than code
  semantics.  From this reference, a complete test suite is derived
  and applied to demonstrate the observational equivalence between the
  synthesised abstract test reference and the generated concrete
  controller code running on a control system platform.
\end{abstract}

\section{Introduction}
\label{sec:intro}

In verified controller synthesis, world models are used that comprise
all potential behaviours of humans, robots, further equipment, and the
controller to be synthesised.  A world model enables quantitative risk
assessment, for example, by stochastic model checking.  Such a model
describes a range of controller behaviours some of which---when
implemented correctly---guarantee that the overall risk in the actual
world is acceptable, provided that the stochastic assumptions have
been made to the safe side.  The objective of the synthesis step is to
select a \emph{controller behaviour} from this range that meets
requirements given as constraints, for example, to stay within an
acceptable risk bound.  Within such constraints, the synthesis
can optimise further objectives, for example, maximal performance or
minimal cost and risk.
Because of crossing the boundaries between different abstractions,
formalisms, and tools, verified controller synthesis for
safety-critical systems naturally has to be accompanied by rigorous
testing.  Indeed, standards and regulations for safety-critical
systems~(e.g. \cite{ISOTS15066,ISO26262,DO178C,DO330}) require testing
as a key element to obtain certification credit before entry into
service.
Hence, a key methodological aim is to bridge the gap between verified
controller synthesis and the generation of executable code that is being
deployed on a control system platform and integrated into the wider system
to be put into service.

\paragraph{Approach.}
\label{sec:approach}

\begin{figure}[t]
  \centering
\includegraphics[width=\linewidth]{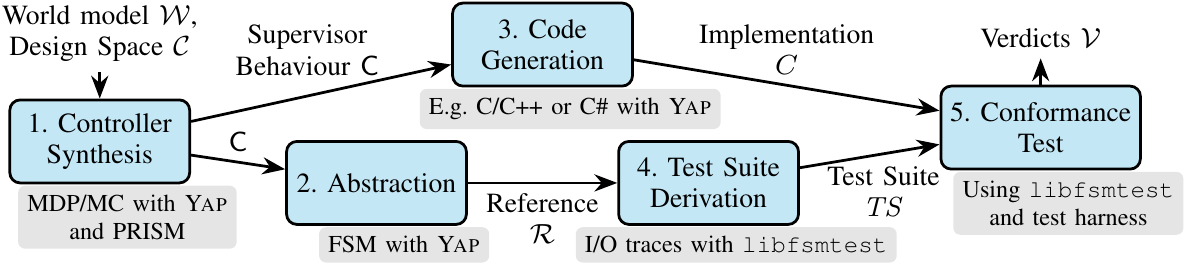}
  \caption{Workflow and artefacts of the proposed tool-supported approach to
    complete testing}
  \label{fig:workflow}
\end{figure}

Following this aim, we propose an \emph{integrated formal approach to
  the complete testing of synthesised supervisory discrete-event
  controllers that enforce safety properties} in domains such as
human-robot collaboration and autonomous driving.  Our  
tool-supported approach works as follows.

\textbf{1. Controller Synthesis.}  The verified synthesis step is based
on policy synthesis for Markov decision processes
\cite{Kwiatkowska2011-PRISM4Verification,
  Kwiatkowska2007-StochasticModelChecking}.  A conceptual \emph{world
  model} $\mathcal{W}$ is constructed that defines all the behaviours of
all the relevant actors~(e.g. humans, robots, other equipment) and the
controller under consideration.  The range of controller behaviours
are denoted as the controller \emph{design space} $\mathcal{C}$.  Then, the
relevant temporal logic properties are formally verified of $\mathcal{C}$ and an
appropriate (optimal) \emph{controller behaviour} $\mathsf{C}\in\mathcal{C}$ is
selected~(synthesised) from~$\mathcal{C}$.  For this step, we adopt the
approach described in \cite{Gleirscher2021-VerifiedSynthesisSafety,
  Gleirscher2020-SafetyControllerSynthesis}.

\textbf{2. Abstraction.}  Then, the selected (verified) controller
behaviour $\mathsf{C}$ is abstracted into a \emph{test reference} model
$\mathcal{R}$.  This model is described as a \emph{symbolic finite state
  machine~(SFSM)}~\cite{DBLP:conf/icst/Petrenko16}, where the control
states are called \emph{risk states}.  Symbols correspond 
    to subsets of $\mathcal{W}$'s state space.  The \emph{input} alphabet
corresponds to the events monitored (observed) by the controller, the
\emph{output} alphabet to the signals that the controller can issue to
$\mathcal{W}$ as the controlled process.  An event is triggered by a
\emph{guard condition}, whose input valuation changes from
\texttt{false} to \texttt{true}, so that a transition labelled with
(or fulfilling) this guard can be taken.  Transitions of $\mathcal{R}$
are labelled with such input/output~(I/O) pairs and derived from
$\mathsf{C}$.

\textbf{3. Code Generation.}  $\mathsf{C}$ is also translated into a
software component $C$ %
executable on the control system platform of a robotic or autonomous
system.  Following an embedded systems tradition, we use C/C++ as the
target language for $C$, making the reasonable assumption that
the used type of FSMs has a simpler semantics than the executable
code.  Abstraction and code generation are explained in
\Cref{sec:deriv-test-refer}.

\textbf{4. Test Suite Derivation.}  Using the
H-Method~\cite{DBLP:conf/forte/DorofeevaEY05}, in this step, a
complete test suite $\mathit{TS}$ for I/O conformance testing is
derived for a finite state machine~(FSM) abstraction of $\mathcal{R}$.
This abstraction maps the SFSM guard conditions to atomic input
labels; otherwise it adopts the SFSM structure without changes. It has
been shown in~\cite{peleska_sttt_2014,Huang2017} that complete FSM
test suites can be mapped to likewise complete suites on SFSMs, when
the FSM input events $e$ are considered as input equivalence classes
of the SFSM, and each $e$ is refined to a concrete SFSM input data
tuple solving the equivalence class constraint (this is just a refined
guard condition).

\textbf{5. Conformance Test.}  Based on a generated test harness
emulating the target platform, the test suite $\mathit{TS}$ is run
against $C$ to record outputs and obtain a complete set of
verdicts $\mathcal{V}$.  A complete pass shown by the verdicts demonstrates the
\emph{observational equivalence} between the test reference $\mathcal{R}$
and the controller code $C$.  Test suite derivation and
conformance test execution are explained in
\Cref{sec:test-strategy}. There, it is also explained how potential
errors in the reference model $\mathcal{R}$, the test suite generator, or
the test harness can be uncovered. This is required according to
standards for safety-critical control applications (see,
e.g.~\cite{DO178C,DO330}), because faulty tool chains might mask
``real'' errors in the software under test.

\paragraph{Related Work.}
\label{sec:related-work}

In the rich body of literature on verified controller synthesis, the
approaches in \cite{Orlandini2013-ControllerSynthesisSafety,
  Bersani2020-PuRSUEspecificationrobotic} from collaborative robotics
are perhaps closest to the one presented here as they include a
platform deployment stage.  While these authors focus on the synthesis of
complete robot controllers, our approach focuses on safety supervisors
but includes a testing step reassuring the correctness of platform
code generation.
The authors of \cite{Villani2019-Integratingmodelchecking} propose a
general integration of quantitative model checking~(with
\textsc{Uppaal}~\cite{Behrmann2004-TutorialUppaal}) with model-based
conformance testing and fault injection.  Apart from using the switch
cover method for test suite generation, their approach is highly
similar to our Mealy-type test reference generation, conformance
testing, and mutation approach for test suite evaluation.  However,
while their focus is more on cross-validation of \textsc{Uppaal} and
FSM models, we concentrate on code robustness tests, assuming that $\mathcal{W}$
has been validated and verified beforehand.

The investigation of complete testing methods is a very active
research field~\cite{Petrenko:2012:MTS:2347096.2347101}.  The
H-Method~\cite{DBLP:conf/forte/DorofeevaEY05} applied for testing in
this paper has been selected because (1) it produces far less test
cases than the ``classical'' W-Method~\cite{chow:wmethod}, but (2) it is
still very intuitive with regard to the test case selection
principles. This facilitates the qualification of the test case
generator, as discussed in Section~\ref{sec:test-strategy}.  If the
main objective of a testing campaign was just to provide complete
suites with a minimal number of test cases, then the
SPYH-Method~\cite{DBLP:conf/icst/SouchaB18} should be preferred to the
H-Method.

Whereas hazard- or failure-oriented %
testing~\cite{Gleirscher2011-HazardbasedSelection%
  ,Lesage2021-SASSISafetyAnalysis} and requirements falsification
based on negative
scenarios~\cite{Uchitel2002-Negativescenariosimplied%
  ,Gleirscher2014-BehavioralSafetyTechnical%
  ,Stenkova2019-GenericNegativeScenarios} are highly useful if no
complete $\mathcal{R}$ is available or if $\mathcal{R}$ still needs to be
validated and revised, our approach is complete once $\mathcal{R}$ is
successfully validated.  That is, any deviation from $\mathcal{R}$
detectable by these techniques is also uncovered by at least one test
case generated by our approach.  Moreover, our approach is usable to
test controller robustness without a realistic simulator for~$\mathcal{W}$.

\paragraph{Contribution.}

We propose a 
solution to the generation of well-defined
test references used in techniques such as the
H-Method~\cite{DBLP:conf/forte/DorofeevaEY05}.  In particular, we
connect test reference generation with the H-Method to derive complete
test suites and demonstrate that this form of robustness testing
yields a correctness proof of a controller under certain assumptions.
We provide tool support for both these steps.  Our proof of concept
indicates that complete test suites are a feasible and practically
attractive means to verify correctness of implementations of the
considered class of discrete-event control modules.
In \Cref{sec:runn-example-robot}, we explain the safety supervisor
concept by means of an example.  In
\Cref{sec:deriv-test-refer,sec:test-strategy}, we explain code and
test reference generation and test suite derivation.  We add concluding
remarks in \Cref{sec:conclusion}.

\section{The Safety Supervisor Concept with an Illustrative Example}
\label{sec:runn-example-robot}

To illustrate our approach, we reuse our example from the domain of
human-robot collaboration in industrial manufacturing as discussed in
\cite{Gleirscher2020-SafetyControllerSynthesis,
  Gleirscher2021-VerifiedSynthesisSafety}.  In this example, a human
operator collaborates with a robot on a welding and assembly
\emph{task} in a work cell equipped with a spot welder.  This setting
involves several actors performing potentially dangerous
actions~(e.g. robot arm movements, welding steps) and, thus, implies
the reaching of hazardous states~(e.g. operator near the active spot
welder, $HC$, or operator and robot on the work bench, $HRW$).  Such
states need to be either avoided or reacted to in order to prevent
accidents from happening or at least to reduce the likelihood of such
undesired events.

This task of \emph{risk mitigation} is, by design, put under the
responsibility of a supervisory discrete-event controller $C$.
This controller is supposed to enforce probabilistic safety properties
of the kind ``the probability of an accident $a$ is less than
$\mathit{pr}_a$'' or ``hazard $h$ happens less likely than
$\mathit{pr}_h$''.  The underlying conceptual controller behaviour
$\mathsf{C}$ comprises
\begin{inparaenum}[(i)]
\item the detection of critical events,
\item the performance of corresponding \emph{mitigation} actions to
  react to such events and reach a \emph{safe} risk state, and
\item, avoiding a paused task or degraded task performance, the
  execution of \emph{resumption} actions to resolve the event and to
  return to a safe but productive risk state.
\end{inparaenum}
For the sake of brevity, we call $C$ a \emph{safety
  supervisor}.

\section{Derivation of the Software Module and the Test Reference}
\label{sec:deriv-test-refer}

We summarise \cite{Gleirscher2021-VerifiedSynthesisSafety} on how to
obtain the world model $\mathcal{W}$, the controller design space
$\mathcal{C}$, and the controller behaviour $\mathsf{C}$.  Then, we
describe in more detail the generation of the controller software
component $C$ (for deployment) and the abstraction into the test
reference $\mathcal{R}$~(for test suite derivation, see
\Cref{sec:test-strategy}).

\begin{wrapfigure}[12]{r}{6cm}
  \vspace{-1em}
  \footnotesize
\includegraphics[width=\linewidth]{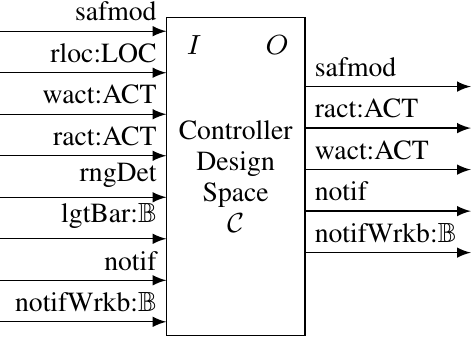}
  \vspace{-1em}
  \caption{Interface between $\mathcal{W}$ and $\mathcal{C}$}
  \label{fig:ctr-interface}
\end{wrapfigure}
The world model $\mathcal{W}$ is a Markov decision process~(MDP), the
result of a fixed-point application of actions given as
probabilistic guarded commands to an initial state of
$\mathcal{W}$~\cite{Kwiatkowska2011-PRISM4Verification}.  MDPs are models
containing \emph{uncertainties} about aspects not under control (or
agency) or not to be modelled explicitly. The world state space is
defined using a set $V$ of finite-sorted variables.  The MDP
is a labelled transition system where the transition relation encodes
non-deterministic and probabilistic choice in a compound manner and
states are labelled with atomic propositions holding of $V$'s
valuations defining the states.  Non-deterministic decisions encode
freedom of choice of the actors in $\mathcal{W}$, in particular, the
controller design space $\mathcal{C}$.  This freedom can be resolved by
picking an \emph{appropriate policy}, a choice resolution for each
state, with the result of obtaining a Markov chain~(MC), a labelled
transition system without indeterminacy in the controller (and the
other considered actors).  Policy appropriateness can be thought of as
sub-setting $\mathcal{C}$ and is defined by constraints to be specified in
probabilistic computation tree
logic~\cite{Kwiatkowska2011-PRISM4Verification}.  The resulting MC is
verified against these constraints and includes the selected behaviour
$\mathsf{C}\in\mathcal{C}$.

\begin{wrapfigure}[14]{r}{6cm}
  \vspace{-1em}
  \footnotesize
\includegraphics[width=\linewidth]{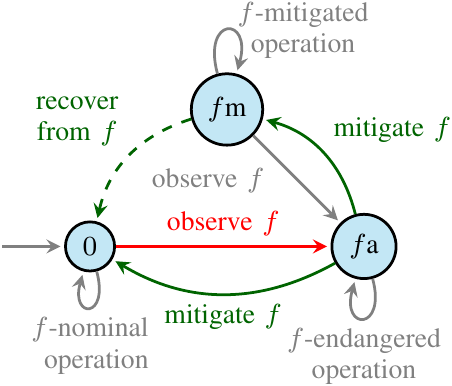}
  \vspace{-1em}
  \caption{Phase transitions of factor $f$}
  \label{fig:factor}
\end{wrapfigure}
Now, $\mathsf{C}$ has to be translated into the two forms
$C$ and
$\mathcal{R}$.  For this step, we define the variables
$I\subseteq V$ to be monitored and the variables
$O\subseteq V$ that can be controlled, resulting in
what we call the \emph{syntactic interface} (alphabet)
$\Sigma\subseteq\type I\times\type O$\footnote{$\type S$ of a set
  $S$ of sorted variables returns the set of tuples in the Cartesian product
  of the sorts of the variables in
  $S$.}   of
$\mathcal{C}$~(see \Cref{fig:ctr-interface}) \cite{Broy2010-LogicalBasisComponent}.
This interface defines the nature of the changes in
$\mathcal{W}$ that any $\mathsf{C}\in\mathcal{C}$ can observe and perform.

The control states of both $C$ and
$\mathcal{R}$ are derived from the notion of \emph{risk states}
\cite{Gleirscher2021-RiskStructuresDesign}, which is defined over a
set $F\subset V$ of
$\mathbb{P}$-sorted variables modelling the critical events considered
in~$\mathcal{W}$ as \emph{risk factors}.  We require
$(I\cup O)\cap F=\emptyset$.  The
sort~$\mathbb{P}=\{0,a,m\}$ models life-cycle stages for handling a
factor
$f\in F$~(\Cref{fig:factor}), for example, from \emph{inactive}
($0$), \emph{active} ($f$a), and \emph{mitigated}
($f$m) back to \emph{inactive}
\cite{Gleirscher2021-RiskStructuresDesign}.  In the example in
\Cref{sec:runn-example-robot}, we consider three factors, hence
$F = \{\mathit{HS}, \mathit{HC},
\mathit{HRW}\}$.  Each
$\mathsf{C}$ can then be associated with a control state space
$S\subseteq\mathbb{P}^{|F|}$.

We then translate the controller fragment $\mathsf{C}$ of the MC
transition relation~(resulting from policy synthesis over $\mathcal{W}$)
into C++ code.  Basically, every transition of $\mathsf{C}$ is translated
into a guarded action
$[a] i\land r\colon (o,r')\leftarrow\textsc{STEP}(i,r)$ with
$r,r'\in S, (i,o)\in\Sigma$ and action name $a$ derived
from $F$, $r$, and $r'$.  For that, the source state of each
transition is mapped into two parts: one corresponding to the input
$i$~(the observed event) and one corresponding to a risk state $r$.
The control and state updates $o$ and $r'$ to be associated with this
action are derived from the difference in the controlled variables
$O\cap F$ between source and target states.  $C$
implements \Cref{alg:codepattern}, intentionally simple (not using
action names) and wrapped into platform-specific code (not shown) for
data processing and communication.

\begin{wrapfigure}[7]{r}{6cm}
  \vspace{-1em}
  \begin{minipage}{6cm}
    \begin{algorithm}[H]
      \footnotesize
      \caption{Safety supervisor} 
      \label{alg:codepattern}
      \begin{algorithmic}[1]
        \Procedure{Ctr}{in \textsf{Event} $i$, out
          \textsf{Signal} $o$}
        \State $r \gets$ \Call{init}{}
        \Comment{init. control/risk state}
        \While{true}
        \Comment{should implement $T$}
        \State 
        $[]_{i\in I,r\in S}\;i\land r\colon (o,r') \gets$
        \Call{step}{$i, r$}
        \label{alg:handle-event}
        \EndWhile
        \EndProcedure
      \end{algorithmic}
    \end{algorithm}
  \end{minipage}
\end{wrapfigure}
\begin{figure}[t]
  \centering
  \includegraphics[width=\textwidth]{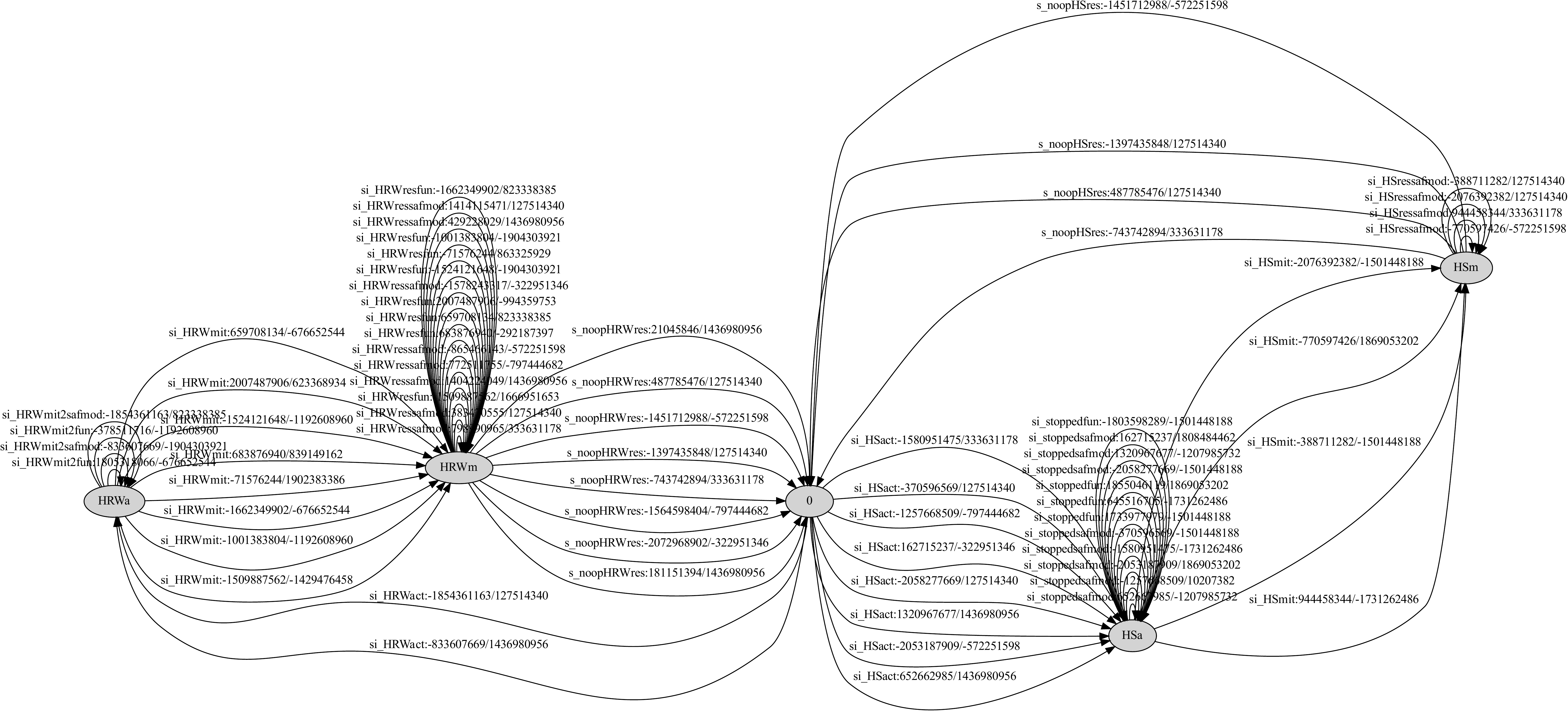}
  \caption{Visual representation of $\mathcal{R}$. Nodes are risk states
    in $S$, edges transitions in $T$.  Because valuation expressions
    for $(i,o)\in\Sigma$ are too long, edge labels are
    hashed and prefixed with action names for readability, following
    the pattern \texttt{$a$:$h(i)$/$h(o)$} with action name $a$ and
    an integer-valued hash function~$h$.}
  \label{fig:testref}
\end{figure}

In order to obtain $\mathcal{R}$, we then translate the
$\mathsf{C}$-fragment of this transition relation into a deterministic
Mealy-type FSM $\mathcal{R} = (S,\Sigma,T,\overline{s})$
with the state space $S$, the transition relation
$T\subseteq S\times\Sigma\times S$, and the
initial state $\overline{s}\in S$ being congruent with the one in $\mathcal{W}$.
\Cref{fig:testref} shows $\mathcal{R}$ for the example in
\Cref{sec:runn-example-robot} and is an operation
refinement %
of the composition of the factors~(\Cref{fig:factor}) of $F$.
The translations into $C$ and $\mathcal{R}$ including the
generation of the test harness are carried through with the \textsc{Yap}\xspace
tool.\footnote{The discussed features are available in \textsc{Yap}\xspace version
  0.8+, see \url{https://yap.gleirscher.at}.}

Regarding the difference of $C$ and $\mathcal{R}$, the semantics
of $C$ can vary significantly.  $\mathcal{R}$ is converted into an
input format suitable for test suite generation via
\texttt{libfsmtest}\xspace~\cite{libfsmtest}.  Here, we consider a C++ component for
a low-level real-time implementation, for example, an FPGA synthesised
from VHDL or Verilog HDL\footnote{Field-programmable gate array (FPGA);
  VHSIC or Verilog hardware description language (VHDL or Verilog HDL)} generated from
C++. In \cite{Gleirscher2021-VerifiedSynthesisSafety}, we consider a
C\# component used in a simulation of $\mathcal{W}$ in a ``Robot Operating
System''-enabled digital twinning environment.  While the semantics of
the C++ and C\# implementations $C_{\mathrm{C++}}$ and
$C_{\mathrm{C\#}}$ may dramatically differ, $\mathcal{R}$ can be
shared between the two.  The only difference on the testing side is in
the mappings used in the test harness~(\Cref{sec:test-strategy}) to
deliver the inputs to $C$ and record the outputs of
$C$.

\section{Complete Test Strategy}
\label{sec:test-strategy}

In this section, we briefly describe the characteristics of complete
test suites, sketch their derivation, outline the chosen recipe for
test suite derivation~(\Cref{sec:strategy}), and discuss typical error
possibilities to be taken into account during standards-oriented
controller and tool certification (\Cref{sec:tool-chain-quali}).

\subsection{Strategy Application}
\label{sec:strategy}

A test suite is complete if---under certain hypotheses---it
guarantees that non-conforming implementations will fail in at least
one test case, while equivalent implementations will always pass the
suite~\cite{Huang2017}. Here, these hypotheses are (a) the number of 
control states implemented in $C$, and (b) assumptions about potential mutations of 
guards and output assignments.
Since the verification of safety-critical systems requires
code to be open source for analyses, these hypotheses can be checked
using static analysis.  We check that $C$ has the same number of control states 
as $\mathcal{R}$, and it is checked that the guard conditions in $\mathcal{R}$ have been correctly translated to corresponding 
branching conditions in $C$~(cf. \Cref{alg:handle-event} in \Cref{alg:codepattern}).

Since the reference   $\mathcal{R}$ is modelled as an SFSM, we need a method to construct 
complete test suites for SFSMs. We follow the recipe from~\cite{peleska_sttt_2014,Huang2017} which allows 
us to use test generation strategies for the simpler class of FSMs and translate the
resulting test suite to an SFSM suite as follows: (1) For the SFSM, input equivalence 
classes are calculated.  This is performed by creating all conjuncts of positive and negated SFSM guard conditions 
which have at least one solution. (2) An FSM is created as an   abstraction of $\mathcal{R}$. The input alphabet of this FSM
consists of the identifiers for the input equivalence classes calculated in~(1). Control states, output events, and transition arrows are
directly adopted from the SFSM. (3) For this FSM, a complete test suite is created using the 
H-Method~\cite{DBLP:conf/forte/DorofeevaEY05}. Its test cases consist of input traces, where each input is an  identifier of an SFSM input equivalence class. The expected results are obtained by running this input trace against the FSM.
(4) The FSM test suite is \emph{refined} to an SFSM test suite by calculating concrete input representatives from the constrains specifying the referenced input classes. (5) The concrete SFSM test suite is executed in a \emph{test harness}: this is an executable running the test cases one by one against $C$ and checking its responses against the FSM test oracle.

The theory elaborated in~\cite{peleska_sttt_2014,Huang2017} confirms that the concrete SFSM test suite is also complete, if this holds for the abstract FSM test suite. Since we know that $C$ has the same number of control states and the same guards as 
$\mathcal{R}$, passing the test suite is equivalent to {\it proving} observational equivalence between   $C$ and $\mathcal{R}$.
For tool support, the \texttt{libfsmtest}\xspace library~\cite{libfsmtest} is used which provides an implementation of 
the H-Method and a template for the test harness.

\subsection{Verification of Verification Results}
\label{sec:tool-chain-quali}

For automated verification/testing of safety-critical system
components, applicable standards require a verification that the tool
chain involved does not mask any errors inside $C$.  This
process is usually called \emph{verification of the verification
  results}. We consider the possible errors in the testing environment
one by one. (1) Error in the generation of $\mathcal{R}$: The complete
test suite created as described above characterises $\mathcal{R}$ up to
observational equivalence. By checking if the test suite is compatible
with the computations of $\mathcal{W}$, it is shown that $\mathcal{R}$ is
correct. (2) Error in the testing theory: It has been shown
in~\cite{DBLP:conf/pts/SachtlebenHH019} that methods of similar
complexity as the H-Method can be mechanically verified using a proof
assistant~(e.g. Isabelle/HOL). (3) H-Method implementation error:
Here, we have two options: in~\cite{DBLP:conf/pts/Sachtleben20} it has
been demonstrated that correct algorithms can be generated while
proving a testing theory to be correct. Alternatively, the generated
test suite can be checked automatically for completeness: from the
specification of the test cases required for the H-Method given
in~\cite{DBLP:journals/sqj/HuangOP19}, a checking tool can be derived
which verifies that the generated suite really contains the test cases
required according to the theory. This checking algorithm would be
\emph{orthogonal} to the test generation algorithm.  This means that it
is highly unlikely that test generator and completeness checker could
contain complementary errors masking each other out. (4) Test harness
error: The test harness could execute the suite in a faulty way that
masks an error in $C$. To ensure that this is not the case, we
apply \emph{mutation testing}. Using the \texttt{clang} compiler
functions for static code analysis, a set $\mathcal{M}_{C}$ of mutants of
$C$ is created in an automated way. Then it is checked for each
mutant in $\mathcal{M}_{C}$ if it is uncovered by the test suite, or if it
is semantically equivalent to the original version of $C$.

\section{Conclusions}
\label{sec:discussion}
\label{sec:conclusion}

We outlined an approach to the complete testing of synthesised
discrete-event controllers that enforce safety properties in
applications such as human-robot collaboration and autonomous driving.
Our aim is to bridge the gap between verified controller synthesis and
certified deployment of executable code.
We illustrate our approach with a human-robot collaboration example
where a safety supervisor makes autonomous decisions on when and how
to mitigate hazards and resume normal operation.  We check the
specificity of the test reference $\mathcal{R}$ and the strength of the
corresponding test suite $\mathit{TS}$ by mutation of the generated
code $C$, modulo semantic equivalence over $\mathcal{M}_{C}$.
We contribute a preliminary synthesis-based test strategy that allows
one to show total correctness of $C$ under certain
implementation assumptions.  The presented approach is automated in a
tool chain: \textsc{Yap}\xspace and a stochastic model checker~(e.g. PRISM
\cite{Kwiatkowska2011-PRISM4Verification}) for MDP generation and
verification, \textsc{Yap}\xspace for test reference and code generation, and
\texttt{libfsmtest}\xspace~\cite{libfsmtest} for test suite derivation.
For testing the integrated system (robot, welding machine, safety
supervisor and simulation of human interactions), the approach
presented here is embedded into a more general methodology for
verification and validation of robots and autonomous systems, starting
at the module level considered here, and ending at the level of the
integrated overall system~\cite{eder_kerstin_2021_5203111}.

\bibliographystyle{eptcs}
\bibliography{}
\end{document}